\documentclass[a4paper,12pt]{revtex4}

\usepackage{amsmath}
\usepackage{amssymb}

\usepackage{graphicx}

\newcommand{\cavsus}{\eta}

\begin{document}

\title{Susceptibility Propagation for Constraint Satisfaction Problems}
\author{Saburo Higuchi$^{1,2}$ and Marc M\'{e}zard${}^{1}$}
\affiliation{
${}^1$Laboratoire de Physique Th\'eorique et Mod\`eles Statistiques,
Universit\'e Paris-Sud, B\^{a}t 100, 91405 Orsay Cedex\\
${}^2$Department of Applied Mathematics and Informatics,
Ryukoku University, Otsu, Shiga, 520-2194, Japan}

\begin{abstract}
We study the susceptibility propagation, a message-passing algorithm to compute correlation functions. It is applied to constraint satisfaction problems and its accuracy is examined.
As a heuristic method to find  a satisfying assignment,
we propose susceptibility-guided decimation where correlations among the variables play an important role.
We apply this novel decimation to locked occupation problems, a class of hard constraint satisfaction problems exhibited recently. It is shown that the present method performs better than the standard belief-guided decimation.
\end{abstract}

\maketitle

\section{Introduction}
\label{sec:introduction}
Message-passing algorithms have shown to be effective in helping to
find solutions of some hard constraint satisfaction problems (CSPs)
like $K$-satisfiability and coloring. The simplest application
consists in using belief propagation (BP), when it converges, in order
to get some estimate of the marginals of each of the variables. One
must then  exploit the information obtained in this way
(which is in general only approximate in a CSP described by a loopy
factor graph).  So far only two methods have been explored thoroughly:
decimation \cite{mezard2002rks,krzakala2007gsa} and reinforcement \cite{chavas2005spd}. Decimation consists in identifying from
some criterion the most ``polarized'' variable (e.g. the one with the
smallest entropy), and in fixing it to its most probable value. After
this variable has been fixed, one obtains a new, smaller, CSP, to
which one can apply recursively the whole procedure (BP followed by
identifying and fixing the most polarized variable). In reinforcement,
one finds from the BP marginals the most probable value of each
variable, and one adds, in the local measure of each variable, an
extra bias in this preferred direction. The new CSP therefore has the
same number of variables as the original one, but the local measure on
each variable has been changed. One iterates this reinforcement
procedure until the variables are infinitely polarized. If the
algorithm is successful this returns a configuration of variables
which satisfies all constraints. These two procedures, BP+decimation
and BP+reinforcement, are remarkably efficient in random CSPs
like $K$-satisfiability \cite{krzakala2007gsa}, graph colouring \cite{krzakala2007gsa}, and
perceptron learning \cite{MezMora}. When one approaches the SAT-UNSAT
threshold of these problems, a more elaborate version which uses the
information on marginals from survey propagation (SP) is more
effective\cite{mezard2002rks,MPZ,chavas2005spd}, and at present the SP-based decimation and reinforcement
methods are the most efficient incomplete SAT solvers for random
$3$-satisfiability.

Recently, a class of problems has been described\cite{ZM}\cite{ZM2} where these
procedures are much less efficient.  These are the locked occupation
problems(LOPs), a class of CSPs where the set of solution consists of
isolated configurations, far away from each other. Apart from the
XORSAT problem\cite{mezard2003tsd} which can be solved by Gaussian elimination, the
random LOPs are very hard to solve in a broad region of the density of
constraints, below their SAT-UNSAT transition. 
For these LOPs, it is known that SP is equivalent with BP.
The BP+decimation
method has been found to give rather poor results, and the
BP+reinforcement, which works better, is still rather limited. One
reason for this hardness is the fact that local marginals often convey
little information on the solution. This has motivated us to explore
some extensions of the message-passing approaches, in which one uses,
on top of local marginals, some correlation properties of the
variables. Several  possibilities to obtain
information on the correlations from message-passing procedures
have been explored recently \cite{montanari2005clc,chertkov_loop,ParisiSlanina,MezMora}.  Here we use the susceptibility propagation initially introduced
in \cite{MezMora}. 
We  show that some of the hard LOPs that could not be solved by
previous methods can now be solved by a mixture of the single-variable
decimation with a new pair-decimation procedure which makes use of 
the knowledge of correlation. 
In the case of binary
variables which we study here, this new procedure amounts to
identifying a strongly correlated pair of variables, and fixing the
relative orientation of the two variables.

The paper is organised as follows.
In Section~\ref{sec:chip}, we introduce the susceptibility propagation, derived as a linear response to belief propagation. 
This method is examined analytically in Section~\ref{sec:properties},
where it is applied to simple systems for which exact fixed points of the iteration are determined.
In Section~\ref{sec:numerical}, it is applied numerically to locked occupation problems and the accuracy of the method is examined: we measures the performance of the decimation process which makes use of the correlations obtained with this method. The final Section~\ref{sec:conclusion} is devoted to conclusion and discussions.

\section{Susceptibility Propagation}
\label{sec:chip}
\subsection{Occupation Problems}
Let us consider an occupation problem,  which consists of 
$|V|=N$ binary variables $x_i\in\{0,1\}$ $(i\in V)$ and$|F|=M$ constraints $\psi_a(x_{i_{a,1}},\ldots,x_{i_{a,k}})=1$ $(a\in F)$. Each constraint involves exactly $k$ variables and is parameterized by  a $(k+1)$-component ``constraint-vector''  $A=(A(0),\ldots,A(k))$ with binary entries defined as follows.
We say a variable $x_i$ is occupied if $x_i=1$. Let $r_a=\sum_{i\in\partial a} x_i$ be the number of occupied variables that are involved in
the constraint $\psi_a$.
By definition, the constraint  $a$ is satisfied ($\psi_a=1$) if and only if $A(r_a)=1$.

An occupation problem is \textsl{locked} if the following three conditions are met\cite{ZM}\cite{ZM2}\cite{zdeborova2008sph}
\begin{itemize}
\item $A(0)=A(k)=0$.
\item $A(r)A(r+1)=0$ for $r=0,\ldots,k-1$.
\item Each variable appears in at least two constraints.
\end{itemize}

Standard examples of locked occupation problems include
positive 1-in-$K$ satisfiability \cite{raymond2007pds} and parity checks \cite{gallager1962ldp}. 

As can be done for general constraint satisfaction problem, 
a factor graph $G=(V,F;E)$ can be associated with an instance of the occupation problems\cite{kschischang2001fga}.
The set of vertices of this bipartite graph $G$ is $V$ and $F$ while the set of edges is $E=\{(i,a)| i\in V,a\in F, \text{ $x_i$ is involved in $\psi_a$}\}$. 
The notion of neighborhood is naturally introduced: 
$\partial a=\{i\in F|(i,a)\in E\}$,
$\partial i=\{a\in V|(i,a)\in E\}$.
For a collection of variables in $S\subset V$, 
we shall write $\underline{x}_S=\{x_i| i \in S\}$. We also use the short-hand notation $\underline{x}=\underline{x}_V$.

\subsection{Belief Propagation Update Rules}
Consider an occupation problem described by a factor graph $G=(V,F,E)$ and a constraint-vector $A$.
For later use, we introduce local `external fields' $h_\ell^x$  $(x\in\{0,1\}, \ell\in V)$, which will be sent to zero at the end, 
and consider a joint probability distribution
\begin{equation}
  p(\underline{x}|\underline{h}^x)
=\frac{1}{Z(\underline{h}^x)}
\prod_{a=1}^M \psi_a(\underline{x}_{\partial a})
\times
\prod_{\ell=1}^N \prod_x \mathrm{e}^{h_\ell^x\delta_{x_\ell,x}} \ .
\label{eq:joint-prob}
\end{equation}
This probability distribution is well defined as soon as there exists at least one (``SAT'') configuration satisfying all the constraints. The constant $Z(\underline{h}^x)$ is a normalization factor. Our final aim is to extract solutions from the uniform measure $ p(\underline{x}|\underline{0}) $ over solutions
satisfying all constraints (when there exists at least one solution).

The marginal distribution $p_i(x_i|\underline{h})$ can be estimated by the BP algorithm.
The BP  update rules for two families of messages, namely cavity fields and cavity biases, are given by \cite{yedidia2003ubp,mezard2009ipa}
\begin{align}
  \nu_{i\rightarrow a}^{(t+1)}(x_i|\underline{h}^x)=&\frac{1}{Z^{(t)}_{i\rightarrow a}(\underline{h}^x)}
\prod_{b\in \partial i\setminus a}
\hat{\nu}_{b\rightarrow i}^{(t)}(x_i|\underline{h}^x)
\times  \prod_x \mathrm{e}^{h_i^x\delta_{x_i,x}},
\label{eq:bpu1}
\\
  \hat{\nu}_{a\rightarrow i}^{(t)}(x_i|\underline{h}^x)=&
\sum_{\underline{x}'_{\partial a}}\delta_{x_i,x'_i}\psi_a(\underline{x}'_{\partial a})
 \prod_{\ell\in\partial a\setminus i} \nu_{\ell\rightarrow a}^{(t)}(x'_\ell|\underline{h}^x).
\label{eq:bpu2}
  \end{align}
Here, we have decided to introduce a normalization factor $Z^{(t)}_{i\rightarrow a}(\underline{h}^x)$ for $\nu_{i\rightarrow a}^{(t)}(x_i|\underline{h}^x)$ and to avoid the normalization for $\hat{\nu}_{a\rightarrow i}^{(t)}(x_i|\underline{h}^x)$. This choice is perfectly valid for BP, and it helps to get relatively simple susceptibility propagation update rules \eqref{eq:chipu1}\eqref{eq:chipu2}.

Assuming convergence to a fixed point, the BP estimate for the marginal distribution of variable $i$ is:
\begin{equation}
  p_i(x_i|\underline{h}^x)=\frac{1}{Z_i(\underline{h}^x)}\prod_{b\in \partial i} \hat{\nu}^{(*)}_{b\rightarrow i}(x_i|\underline{h}^x),
\label{eq:marginal}
\end{equation}
where $ \hat{\nu}_{a\rightarrow i}^{(*)}(x_i|\underline{h}^x)$  is the fixed point  of the BP iteration.

\subsection{Susceptibility Propagation Update Rules}
The 2-point connected correlation function at $\underline{h}=\underline{0}$ is obtained as
\begin{equation}
  p_{ij}^\mathrm{conn}(x_i,x_j)\equiv p_{ij}(x_i,x_j)-p_i(x_i)p_j(x_j)=\left.\frac{\partial p_i(x_i|\underline{h}^x)}{\partial h_j^{x_j}}\right|_{\underline{h}=\underline{0}}.
\end{equation}
To have a message-passing algorithm to calculate this quantity, 
we introduce the cavity susceptibility and its companion by
\begin{equation}
  \nu_{i\rightarrow a,j}(x_i,x_j)=
\left.  \frac{\partial\nu_{i\rightarrow a}(x_i|\underline{h}^x)}{\partial h_{j}^{x_j}}\right|_{\underline{h}=\underline{0}},
\end{equation}
\begin{equation}
  \hat{\nu}_{a\rightarrow i,j}(x_i,x_j)=
\left.  \frac{\partial\hat{\nu}_{a\rightarrow i}(x_i|\underline{h}^x)}{\partial h_{j}^{x_j}}\right|_{\underline{h}=\underline{0}}.
\end{equation}
Note that the roles of variables $x_i$ and $x_j$ are asymmetric; $j$ can be an arbitrary variable while $i$ is a neighbor of the constraint $a$.

The cavity susceptibility and its companion
can be calculated by a message-passing method \cite{montanari2005clc}.
The susceptibility propagation update rules can be obtained by differentiating the belief propagation update rules \eqref{eq:bpu1} and \eqref{eq:bpu2} with respect to $h_j^x$. They read\cite{MezMora}\cite{Mora2007}
\begin{align}
  \nu_{i\rightarrow a,j}^{(t+1)}(x_i,x_j)=&
\frac{1}{Z^{(t)}_{i\rightarrow a}(\underline{h}^x)}\prod_{b\in \partial i\setminus a}
\left(\delta_{i,j}\delta_{x_i,x_j}
+\sum_{b\in\partial i\setminus a}
\frac{\hat{\nu}^{(t)}_{b\rightarrow i,j}(x_i,x_j)}{\hat{\nu}_{b\rightarrow i}^{(t)}(x_i)}
+C_{i\rightarrow a,j}^{(t)}(x_j)
\right), 
\label{eq:chipu1}\\
  \hat{\nu}_{a\rightarrow i,j}^{(t)}(x_i,x_j)=&
\sum_{\underline{x}' _{\partial a}}
\delta_{x_i',x_i'}
\psi_a(\underline{x}_{\partial a}')
\times
\left(
\prod_{\ell\in\partial a\setminus i}
\nu_{\ell\rightarrow a}^{(t)}(x_\ell')
\right)
\sum_{m\in\partial a\setminus i}
\frac{{\nu}^{(t)}_{m\rightarrow a,j}(x_m',x_j)}{{\nu}_{m\rightarrow a}^{(t)}(x_m')},
 \label{eq:chipu2}
\end{align}
where
\begin{equation}
  \nu_{i\rightarrow a}^{(t)}(x_i)  
=  \nu_{i\rightarrow a}^{(t)}(x_i|\underline{h}^x=\underline{0}),\quad\quad
  \hat{\nu}_{a\rightarrow i}^{(t)}(x_i)
=  \hat{\nu}_{a\rightarrow i}^{(t)}(x_i|\underline{h}^x=\underline{0}).
\end{equation}
The function $C_{i\rightarrow a,j}^{(t)}(x_j)$ originates from the derivative of $Z^{(t)}_{i\rightarrow a}$ and can be determined by requiring the normalization
\begin{equation}
\sum_{x_i} \nu^{(t)}_{i\rightarrow a,j}(x_i,x_j)=0.
\end{equation}


Let us suppose that we have found a fixed point of BP and the susceptibility propagation. By differentiating \eqref{eq:marginal} with respect to the external fields,
we can express the 2-point connected correlation function in terms of the messages at the fixed point as
\begin{equation}
p_{ij}^\mathrm{conn}(x_i,x_j)=p_i(x_i)[\delta_{i,j}\delta_{x_i,x_j} + C_{ij}(x_j)]
+\frac{1}{Z_i(\underline{0})}\sum_{b\in\partial i} \hat{\nu}_{b\rightarrow i,j}^{(*)}(x_i,x_j)
\prod_{c\in\partial i\setminus b}\hat{\nu}_{c\rightarrow i}^{(*)}(x_i).
\label{eq:connected-correlation}
\end{equation}
The constant $C_{ij}(x_j)$ is related to the derivative of $Z_i(\underline{h})$ and is conveniently fixed by the condition $\sum_{x_j} p_{ij}^\mathrm{conn}(x_i,x_j)=0$.

\subsection{Log-likelihood representation}
The rules (\ref{eq:chipu1},\ref{eq:chipu2}) apply to all types of CSPs with discrete variables. 
When dealing with binary variables, it is helpful to rewrite the belief and susceptibility update equations in
terms of log-likelihood variables.
We introduce the cavity field and cavity bias in the log-likelihood representation $n_{i\rightarrow a}$ and ${\hat n}_{a\rightarrow i}$ as
(we omit the time superscript $(t)$ where it is obvious):
\begin{align}
  \nu_{i\rightarrow a}(x_i|\underline{h})=&A_{i\rightarrow a}\; \mathrm{e}^{n_{i\rightarrow a}(\underline{h})s_i},\\
  \hat{\nu}_{a\rightarrow i}(x_i|\underline{h})=&B_{a\rightarrow i}\; \mathrm{e}^{\hat{n}_{a\rightarrow i}(\underline{h})s_i},
\end{align}
where $s_i$ is the spin variable $s_i=2x_i-1=\pm1$ and the external fields in the two representations are related by
\begin{equation*}
  h_j=\frac{h_j^1-h_j^0}{2}.
\end{equation*}


Naturally we define the cavity susceptibility in the log-likelihood representation as
\begin{align}
  \cavsus_{i\rightarrow a,j}=\left.\frac{\partial n_{i\rightarrow a}(\underline{h})}{\partial h_j}\right|_{\underline{h}=0}
\ \ , \ \ 
  \hat{\cavsus}_{a\rightarrow i,j}=&
\left.\frac{\partial \hat{n}_{a\rightarrow i}(\underline{h})}{\partial h_j}\right|_{\underline{h}=0}
\end{align}

The belief propagation update rules read
\begin{align}
  n_{i\rightarrow a}^{(t+1)}=&\sum_{b\in\partial i\setminus a} \hat{n}_{b\rightarrow i}^{(t)} + h_i,
\label{eq:bpu1_log}
\\
  \hat{n}_{a\rightarrow i}^{(t)}=&f_{a\rightarrow i}( \{n_{j\rightarrow a}^{(t)}\}_{j\in\partial a\setminus i}),
\label{eq:bpu2_log}
\end{align}
where
\begin{align}
  f_{a\rightarrow i}( \{n_{j\rightarrow a}\}_{j\in\partial a\setminus i})
=&\frac12\log\frac{F(+1)}{F(-1)},\\
F(\sigma)=&\sum_{\underline{s}_{\partial a}} \delta_{s_i,\sigma}\psi_a(\underline{s}_{\partial a})\prod_{j\in\partial a\setminus i}\mathrm{e}^{n_{j\rightarrow a} s_j}.
\end{align}
By differentiating both sides of (\ref{eq:bpu1_log},\ref{eq:bpu2_log}), we obtain
\begin{align}
\cavsus_{i\rightarrow a,j}^{(t+1)}=&\sum_{b\in\partial i\setminus a} \hat{\cavsus}_{b\rightarrow i,j}^{(t)}+\delta_{i,j}
\label{eq:chipu1_log}
\\
\hat{\cavsus}_{a\rightarrow i,j}^{(t)}=&\sum_{m\in\partial a\setminus i} 
\frac{\partial f_{a\rightarrow i}( \{n_{j\rightarrow a}^{(t)}\}_{j\in\partial a\setminus i})}{\partial n_{m\rightarrow a}}
\times \cavsus_{m\rightarrow a,j}^{(t)}.
\label{eq:chipu2_log}
\end{align}
Assuming that a solution $n_{j\rightarrow a}^{(t)}$ of the BP equations (\ref{eq:bpu1_log},\ref{eq:bpu2_log}) is used, one sees that 
the susceptibility propagation update rule (\ref{eq:chipu1_log},\ref{eq:chipu1_log}) is an inhomogeneous  linear system in $\cavsus$ and $\hat{\cavsus}$.
The coefficient matrix takes the following form:
\begin{align}
\frac{\partial f_{a\rightarrow i}( \{n_{j\rightarrow a}\}_{j\in\partial a\setminus i})}{\partial n_{m\rightarrow a}}
=\frac{\langle s_m s_i\rangle-\langle s_m \rangle \langle  s_i\rangle}{1-\langle s_i\rangle^2}
\label{eq:coeffmat-log}
\end{align}
where $i,m\in\partial a$ and 
$\langle \cdot \rangle$ means 
Here $\langle s_i\rangle$ and $\langle s_m s_i\rangle $ for $i,m\in\partial a$ means
the expectation value with respect to the joint probability distribution for variables that are neighbors of a constraint  obtained solely from beliefs\cite[Sec.14.2.3]{mezard2009ipa}.

In the log-likelihood representation, the magnetization and the pair correlation are given  in terms of the fixed-point messages by
\begin{align}
\langle s_i \rangle =&\tanh\left(\sum_{b\in \partial i} \hat{n}_{b\rightarrow i}^{(*)}
\right),\\
\langle s_i s_j\rangle_\mathrm{conn}\equiv
\langle s_is_j \rangle- \langle s_i \rangle\langle s_j \rangle=&\left[1-\tanh^2\left(\sum_{b\in\partial i} \hat{n}_{b\rightarrow i}^{(*)}\right)\right]\times\left[\sum_{c\in\partial i} \hat{\cavsus}_{c\rightarrow i,j}^{(*)} + \delta_{i,j}\right].
\end{align}
In the above expression, $i$ and $j$ can be arbitrary variables on the factor graph.

\section{Properties}
\label{sec:properties}
\subsection{Linear Equation}
\label{sec:linear}
In order to study the structure of susceptibility propagation update rules 
(\ref{eq:chipu1_log},\ref{eq:chipu2_log}),
we construct a $kMN$-component column vector
\begin{equation}
  \mathbf{y}^{(t)}=(\cavsus_{i\rightarrow a,j}^{(t)},\hat{\cavsus}_{a\rightarrow i,j}^{(t)})^{\mathrm{t}}_{(i,a)\in E, j\in V}.
\end{equation}
Then the fixed point condition associated with (\ref{eq:chipu1_log},\ref{eq:chipu2_log}) can be written as a linear equation
\begin{equation}
  \mathbf{y}^{(*)}=\mathbf{M}\mathbf{y}^{(*)}+ \mathbf{b},
\end{equation}
with the inhomogeneous term
\begin{equation}
  \mathbf{b}=(\delta_{i,j},0)^{\mathrm{t}}_{(i,a)\in E, j\in V}
\end{equation}
The coefficient matrix is block-diagonal in $j$:
\begin{gather}
  \mathbf{M}_{(iaj),(i'a'j')}=\delta_{j,j'} \mathcal{M}_{(ia),(i'a')},\\
\mathcal{M}=
\left(
\begin{array}{cc}
0  & \mathbf{1}(a'\in \partial i \setminus a)\delta_{i,i'}   \\
\mathbf{1}(i'\in \partial a\setminus i) \delta_{a,a'}  
\frac{\partial f_{a\rightarrow i}( \{n_{j\rightarrow a}^{(*)}\}_{j\in\partial a\setminus i})}{\partial n_{i'\rightarrow a'}} & 0  \\
\end{array}
\right),
\end{gather}
where the block $\mathcal{M}$ is independent of the block index $j$.

Thus we obtain the unique fixed point
\begin{equation}
  \mathbf{y}^{(*)}=(\mathbf{1}-\mathbf{M})^{-1}\mathbf{b} \label{eq:y-fp}
\end{equation}
if $(\mathbf{1}-\mathbf{M})$ is invertible,
which is equivalent to the invertibility of  $(\mathbf{1}-\mathcal{M})$.

The susceptibility propagation update rules (\ref{eq:chipu1_log},\ref{eq:chipu2_log}) can be regarded as an iterative method to solve the linear equation equation  \eqref{eq:y-fp}. 
It converges to a value irrespective of the initial vector if all the  eigenvalues of $\mathcal{M}$ have moduli smaller than unity. 
Because the block $\mathbf{M}$ does not depend on $j$,
the existence of the fixed points
and convergence to them are solely determined by $\mathcal{M}$ and do not depend on $j$.

\subsection{Application to simple problems}
When the factor graph is a tree,  even in presence of the external fields $\underline{h}^x$,
the exact marginals are obtained by \eqref{eq:marginal} on a fixed point 
$\nu_{i\rightarrow a}^{(*)}, \hat{\nu}_{i\rightarrow a}^{(*)}$ \cite{kschischang2001fga}.
Therefore, by differentiation with respect to $ \underline{h}^x$ , there exists a susceptibility fixed-point which gives the exact 2-point correlation function. In the examples which we have considered, the iteration of susceptibility propagation converges to this fixed-point.
On the other hand, if the graph has more than one loop, there is no guarantee either that the fixed point exists or the iteration leads to that fixed point. In order to test these statements, we have studied a simple problem, 
the 1-in-2 satisfiability problem, or anti-ferromagnetic Ising model. 

We first study this problem on a chain of length $N$.
Namely, we take $k=2$ and $A=(0,1,0)$, and $V=\{0,1,2,\ldots,N-1\}$, $F=\{0+\tfrac12, 1+\tfrac12,\ldots,N-1-\tfrac12\}$.
This gives  a simple case of a tree factor graph
with $E=\{(i,i+\tfrac12)| i=0,1,\ldots,N-2\}\sqcup\{(i,i-\tfrac12)| i=1,\ldots,N-1\}$.


Away from the boundaries,, since $\partial a$ and $\partial i$ consist of only two variables and constraints, respectively, 
\eqref{eq:bpu1_log},
\eqref{eq:bpu2_log},
\eqref{eq:chipu1_log},
\eqref{eq:chipu2_log}
are simplified to yield
\begin{eqnarray}
  n_{i\rightarrow i\pm\tfrac12}^{(t+1)}
&=&\hat{n}_{i\mp\tfrac12\rightarrow i}^{(t)} \label{eq:bpup1_log_chain}\ \ \  ; \ \  \
  \hat{n}_{i\pm\tfrac12 \rightarrow i}^{(t)}
=-n_{i\pm1\rightarrow i\pm\tfrac12}^{(t)},\label{eq:bpup2_log_chain}\\
  \eta_{i\rightarrow i\pm\tfrac12, j}^{(t+1)}
&=&\hat{\eta}_{i\mp\tfrac12 \rightarrow i,j}^{(t)} + \delta_{i,j}\label{eq:chipup1_log_chain} \ \ \ ; \ \  \
  \hat{\eta}_{i\pm\tfrac12\rightarrow i,j}^{(t)}
=-\eta_{i\pm1 \rightarrow i\pm\tfrac12,j}^{(t)}.
\label{eq:chipup2_log_chain} 
\end{eqnarray}
On the boundary, on the other hand, one has
\begin{equation}
  n_{0\rightarrow\tfrac12}=n_{N-1\rightarrow N-\tfrac32}=0,\quad
  \eta_{0\rightarrow\tfrac12,j}=\delta_{j,0}, \quad \eta_{N-1\rightarrow N-\tfrac32,j}=\delta_{j,N-1}.
\end{equation}
This in turn implies that 
\begin{gather}
  n_{i\rightarrow i\pm\tfrac12}^{(*)}=\hat{n}_{i\pm\tfrac12\rightarrow i}^{(*)}=0\ \ \ ; \ \ \
  \cavsus _{i\rightarrow i\pm\tfrac12,j}=
  \begin{cases}
    (-1)^{i-j} & (\pm(i-j)\geq0)\\
    0         & (\text{otherwise})
  \end{cases}
\end{gather}
which gives:
\begin{gather}
  \langle s_i \rangle=0 \ \ , \ \ 
  \langle s_i s_j \rangle_{\mathrm{conn}}=
\left[-n_{i-1\rightarrow i-\tfrac12,j}^{(*)}-n_{i+1\rightarrow i+\tfrac12,j}^{(*)}\right] + \delta_{ij}
=(-1)^{i-j}.
\end{gather}
In summary, for 1-in-2 satisfiability on a chain, which is a simple XORSAT problem \cite{creignou99} with a tree factor graph,
the belief propagation and susceptibility propagation give the correct magnetization and susceptibility.

Consider now the same problem on the simplest  graph with one loop, a ring.

Namely, let $G$ be a 1-dimensional ring, which is defined by
identifying variable $i=0$ with $N$ and
adding a factor $a=N-\tfrac12$ as well as two incident edges $(N,N-\tfrac12)$ and $(N-1,N-\tfrac12)$.
Moreover, we assume that $N$ is an even integer so that there is no frustration.

BP has a continuous family of fixed points:
\begin{equation}
  n_{i\rightarrow i\pm\tfrac12}^{(*)}=\hat{n}_{i\mp\tfrac12\rightarrow i}^{(*)}=(-1)^iA_\pm,
\end{equation}
where $A_\pm$ is a constant \cite{mezard2009ipa}.
As a consequence of the existence of this family of fixed points,
$(1-\mathcal{M})$ is not invertible; 
in fact it has an  eigenvector  with zero eigenvalue, $\mathbf{y}_0=(\mathbf{1},-\mathbf{1})$ where 
$\mathbf{1}$ corresponds to the $\cavsus$-block and $-\mathbf{1}$ corresponds to the $\hat{\cavsus}$ block.
In agreement with the existence of this dangerous eigenvector,  one finds that the susceptibility propagation update rule does not converge. As the susceptibility messages are updated, 
$\cavsus_{i\rightarrow i+\tfrac12,j}$ picks up the constant shift $\delta_{i,j}=1$.
This effect is accumulated as the messages go around the ring, and the consequence is that the messages diverge as $t\rightarrow\infty$.

\begin{figure}
  \centering
    \includegraphics[scale=0.8]{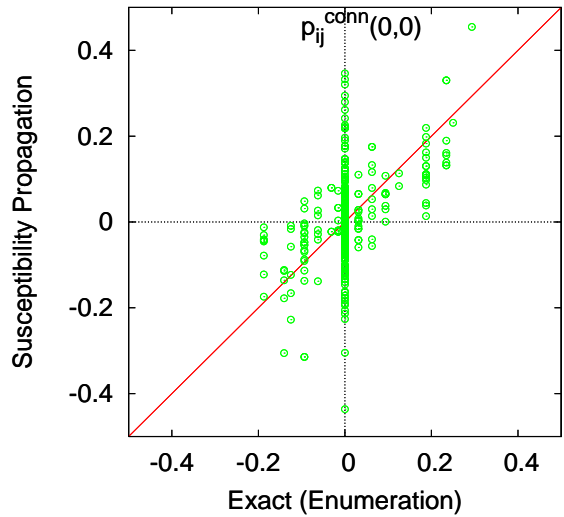}      
    \caption{ \label{fig:exact-sus}
Comparison between the 2-point connected correlation function 
calculated exactly and that estimated with susceptibility propagation. 
A 1-in-4 satisfiability instance on a randomly generated factor graph
with $N=27$ variables and $M=16$ constraints with Poisson degree distribution with average degree $\overline{\ell}=2.4856$.}
\end{figure}

In summary, for 1-in-2 satisfiability on a ring, which is an XORSAT problem on a graph with a loop,
the belief propagation can converge to a family of solutions for the magnetization among which only one solution is exact.
On the other hand, the susceptibility propagation update does not have a fixed point, it diverges. In the simple case of a ring, this behaviour can be cured by using  the finite temperature version of the BP and susceptibility propagation update equations. But in general there is no guarantee of convergence of loopy BP and loopy susceptibility propagation, and when they converge the quality of their results cannot be assessed a priori. Fig.\ref{fig:exact-sus} gives an example of analysis of a small instance of 1-in-4 satisfiability, giving an idea of the errors made by susceptibility propagation on small factor graphs. On the other hand, as for standard BP, one may hope that the method becomes better for large instances when the factor graph is locally tree-like.

\section{Numerical investigation of susceptibility propagation in locked occupation models}
\label{sec:numerical}
In this section we study the use of susceptibility propagation, together with decimation, in some locked occupation models. Specifically, we shall study 
random instances of a locked occupation problem, where
the factor graph is uniformly chosen among the graphs with the following degree distribution. All function nodes have degree $K$ and the variables have random degrees chosen from truncated Poisson degree distribution
\begin{equation}
  q(\ell)=
  \begin{cases}
    0 & (\ell=0,1)\\
    \frac{\mathrm{e}^{-c}c^\ell}{\ell! (1-(1+c)\mathrm{e}^{-c})}. & (\ell\geq2)
  \end{cases},
\end{equation}
for which the average degree is 
\begin{equation}
  \overline{\ell}=\sum_{\ell=0}^\infty \ell q(\ell)=\frac{c(1-\mathrm{e}^{-c})}{(1-(1+c)\mathrm{e}^{-c})}.
\end{equation}

The basic message-passing algorithm that we use is described by the following pseudocode:

\texttt{Input:}
 Factor graph, constraint-vector, convergence criterion, initial messages

\texttt{Output:}
Estimate for 2-point connected correlation functions (or ERROR-NOT-CONVERGED)

  \begin{itemize}
    \item Initialize messages
    \item Repeat until everything converges
      \begin{itemize}
        \item Update cavity fields and cavity biases $\nu_{i\rightarrow a}^{(t)}(x_i)$ and $\hat{\nu}_{a\rightarrow i}^{(t)}(x_i)$ with \eqref{eq:bpu1},\eqref{eq:bpu2}

      \item Update cavity susceptibilities
$\nu_{i\rightarrow a,j}^{(t)}(x_i,x_j)$ and $\hat{\nu}_{a\rightarrow i,j}^{(t)}(x_i,x_j)$ with \eqref{eq:chipu1}\eqref{eq:chipu2}
with the help of $\nu_{i\rightarrow a}^{(t)}(x_i)$ and $\hat{\nu}_{a\rightarrow i}^{(t)}(x_i)$ obtained above

    \end{itemize}
  \item Compute 1-variable marginals $p_i(x_i)$
from the fixed-point messages $\hat{\nu}_{a\rightarrow i}^{(*)}(x_i)$
by \eqref{eq:marginal}
 \item Compute 2-point connected correlation functions  
$p_{ij}^\mathrm{conn}(x_i,x_j)$ 
from the fixed-point messages
$\hat{\nu}_{a\rightarrow i}^{(*)}(x_i)$ and $\hat{\nu}_{a\rightarrow i,a}^{(*)}(x_i,x_j)$
by \eqref{eq:connected-correlation}
\end{itemize}
This algorithm requires a memory proportional to 
$kMN$, and each step of iteration requires a computation of  $\mathcal{O}(N^2)$ for fixed $k$.

\subsection{Decay of correlations}
\label{sec:decay}
\begin{figure}
  \centering
  \includegraphics[scale=0.8]{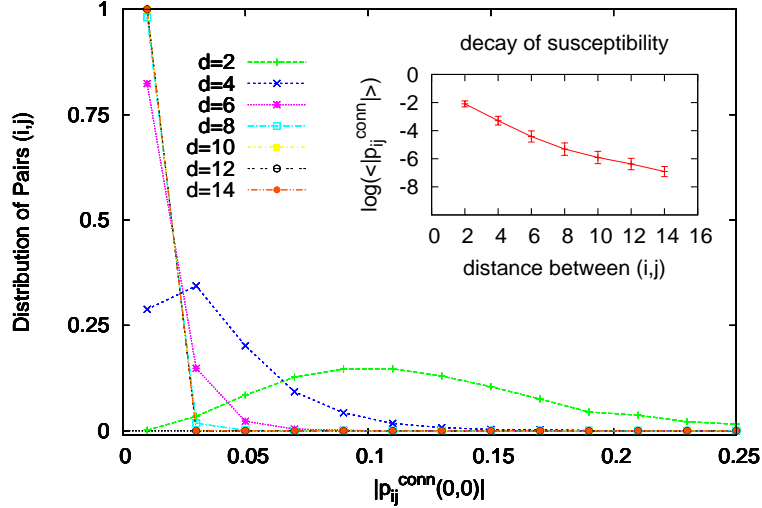}
  \caption{\label{fig:dist-sus}  
    This graph shows how the 2-point connected correlation $p_{ij}^\mathrm{conn}(x_i,x_j)$ decays as the distance $d$ between $x_i$ and $x_j$ increases.
    At each distance $d$, the distribution of $|p_{ij}^\mathrm{conn}(0,0)|$ is plotted. In the inset, logarithm of the average of that quantity is plotted against the distance.
The instance is 1-in-4 Satisfiability 
on a random factor graph with 
$N=1618$ variables and $M=1000$ factors with the truncated Poisson degree distribution with average degree $\overline{\ell}=2.4856$. 
}
\end{figure}

Fig.\ref{fig:dist-sus} shows the distribution of magnitude of 2-point connected correlation function computed with susceptibility propagation for all pairs of points in a graph for a fixed distance between the points. One observes a broad dispersion of correlations, and an approximate exponential decay with the distance.
Here we measure the distance $d$ with the convention that each edge connecting a variable to a constraint is of length 1.
Because of this exponential decay, it is possible to use in some cases  approximate versions of susceptibility propagation which are faster and use less memory. This is done  by truncating to zero 
the cavity susceptibilities $\nu_{i\rightarrow a,j},\hat{\nu}_{a\rightarrow i,j}$
beyond some prescribed distance 
$\mathrm{dist}(a,j)>d$ 
or
$\mathrm{dist}(i,j)>d$ 
and keeping only the correlation functions between pairs of variables not far from each other.
Although one can estimate the 2-variable marginal distribution
solely from the knowledge of cavity fields \cite[Sec.14.2.3]{mezard2009ipa},
this truncation provides us with a more efficient  practical method to compute the 2-variable correlations between variables with $d\geq 4$.

\subsection{Pair Decimation Algorithm}
\begin{figure}
  \centering
 \includegraphics[scale=1.00]{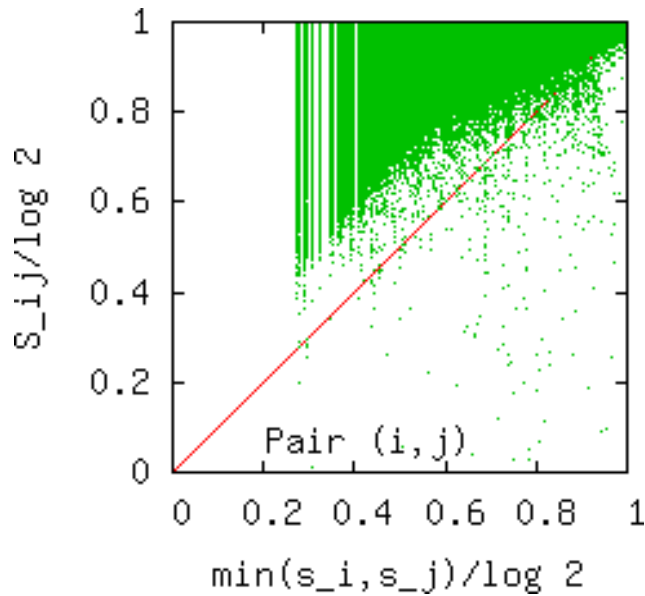}    
 \caption{\label{fig:compare-entropy}
   Comparison between the
 minimum entropy $\min(S_i,S_j)$ (where $S_i$ and $S_j$ are the entropies of $x_i$, $x_j$ )and $S_{ij}$, that of $y_{ij}=\mathbf{1}(x_i=x_j)$.
The instance is 1-in-4 Satisfiability on a random factor graph with 
$N=1618$ variables and $M=1000$ factors with the truncated Poisson degree distribution with average degree $\overline{\ell}=2.4856$. 
}
\end{figure}

As we mentioned in the introduction, 
decimation consists in finding a variable with the smallest entropy and fixing it to the most probable value.
Assuming that the susceptibility propagation provides us with the good estimate for the 2-point connected correlation, we can think of decimation which acts on  a pair of variable instead of a single variable.
Let $x_i$ and $x_j$ be variables. 
If one defines a 
random  variable $y_{ij}=\mathbf{1}(x_i=x_j)$, 
one can compute the entropy for $y_{ij}$ once one knows the 2-variable marginal
$p_{ij}(x_i,x_j)=p_i(x_i)p_j(x_j)+p_{ij}^{\mathrm{conn}}(x_i,x_j)$.
In pair decimation, one identifies the pair $(i,j)$ with the smallest entropy of $y_{ij}$ and one fixes 
 either $x_i=x_j$ or $x_i+x_j=1$, depending on which event is the most probable according to the measured correlation. This results in a reduced smaller CSP, which is still an occupation problem.
The efficiency of this novel decimation process depends on 
whether we can find a pair with less entropy than 
the single variable with the smallest entropy.
It is easy to see that, in the absence of correlations, namely if 
$p_{ij}(x_i,x_j)=p_i(x_i)p_j(x_j)$, 
then the entropy of $y_{ij}$ is larger than the one of $x_i$ or $x_j$. So the whole procedure relies on being able to detect correlations. 
Fig.\ref{fig:compare-entropy} shows that strongly correlated pairs can be found.

In practice, we have used the following decimation algorithm which mixes the two strategies of single-variable decimation and pair decimation:

\texttt{Input:}
 Factor graph, constraint-vector, convergence criterion, initial messages

\texttt{Output:}
A satisfying assignment (or FAIL-NOT-FOUND)

\begin{itemize}
\item While graph has more than $R$ variables:
  \begin{itemize}
\item Compute local entropy estimates for the 1-variable marginals
\item Compute local entropy estimates for the 2-variable marginals
\item if `heuristic criterion finds that single-variable decimation is better',
  \begin{itemize}
  \item then fix the value of the variable.
  \item else identify a variable in the pair with the other (or its negation)
  \end{itemize}
\item Locate completely locked nearest neighbor pairs
\item Clean the graph 
  \begin{itemize}
    \item Fix the value of isolated variables
  \end{itemize}
\item Do warning propagation.
\item Identify local locked pairs
\end{itemize}
\item When the number of variables is equal to or smaller than $R$:  perform an exhaustive search for satisfying assignments. If found
  \begin{itemize}
  \item Then return the satisfying assignment
  \item Else return FAIL-NOT-FOUND
  \end{itemize}
\end{itemize}
The heuristic criterion that we use in order to decide between the two types of decimation is the following. 
We locate a variable with the least entropy
and a pair of variables with the least entropy for $y_{ij}$.
When the former is less than $S_\mathrm{th}$ or is smaller than the latter, we choose to do single-variable decimation.

For the optimal reduction of the entropy within a decimation step, 
it is reasonable to set $S_\mathrm{th}=0$. However, we find that $S_\mathrm{th}>0$ performs better for finding a satisfying assignment.
The optimal value of $S_\mathrm{th}$ depends on the type of locked occupation model and the average degree.
This fact can be interpreted as follows:
the estimation of 1-variable marginals is more precise than the 2-variable ones within given computational resource, 
thus it is advantageous to respect the former if it is decisively small.

Warning propagation is a message-passing algorithm described in \cite{mezard2003tsd,montanari2007scs}. It logically infers the value of variables one by one from local structure of the factor graph.

In the identification of local locked pairs, we look at each degree-2 constraint and see if the constraint enforces $y_{ij}=0$ or $y_{ij}=1$. If it is the case, we identify this pair.

The threshold for exhaustive search has been fixed in our simulations to $R=16$.
\begin{figure}
  \centering
  \includegraphics[scale=0.6]{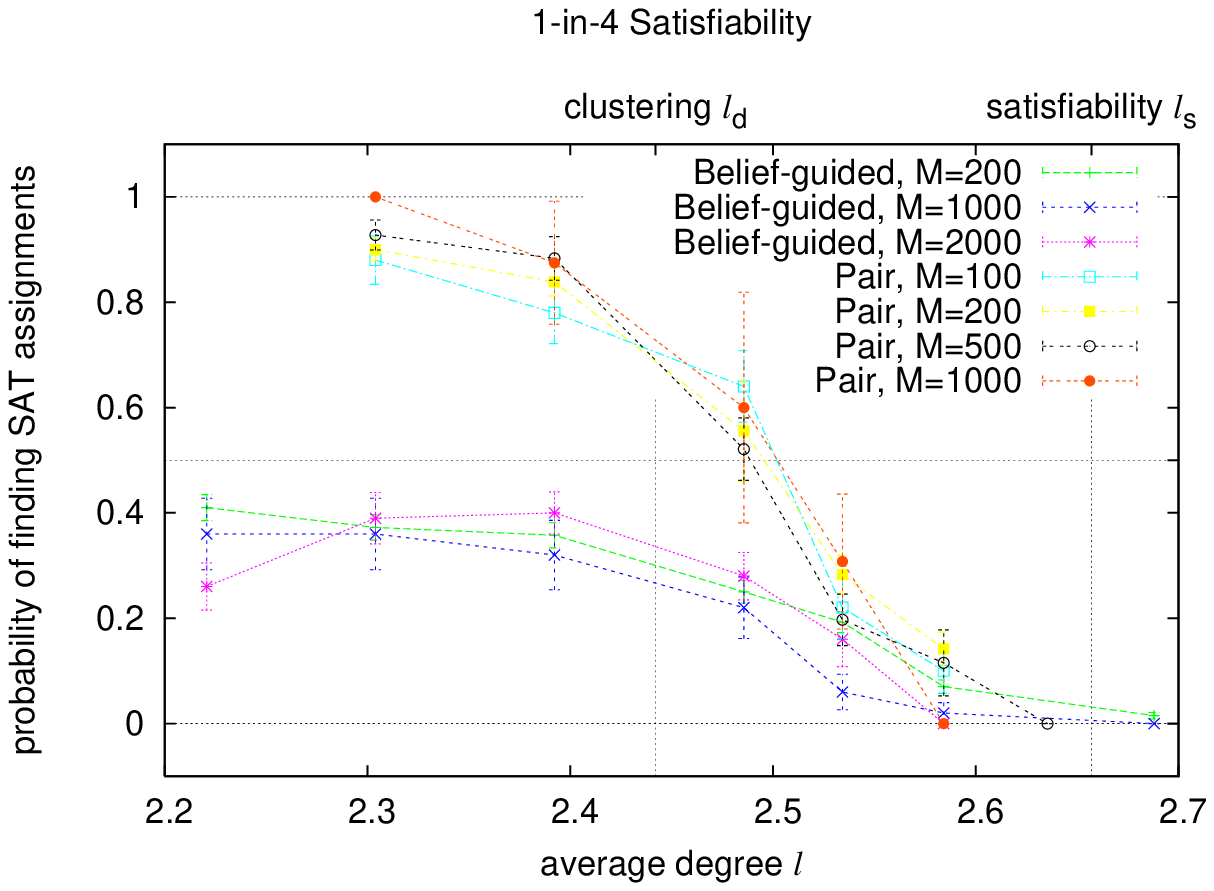}
  \includegraphics[scale=0.6]{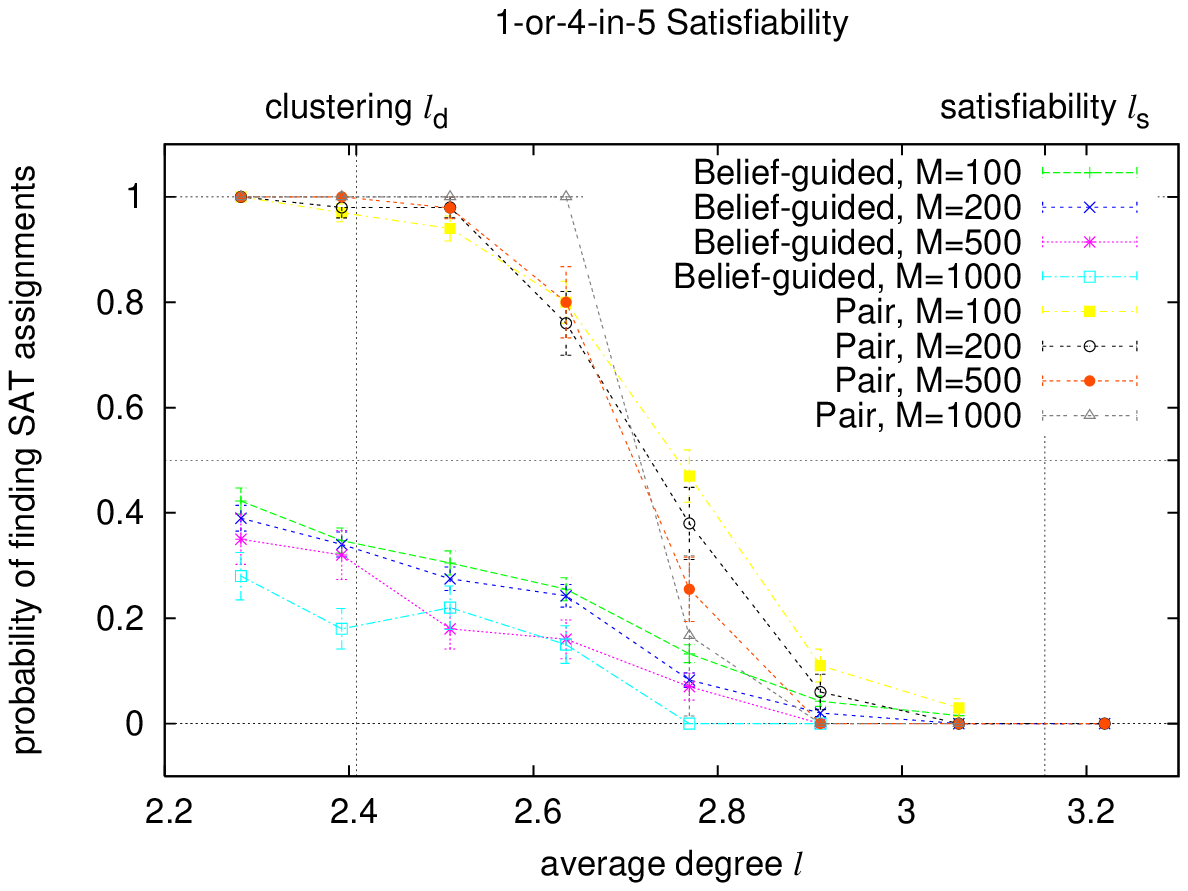}
  \caption{\label{fig:decimation}Success probability of pair decimation process for 1-in-4 satisfiability $A=(0,1,0,0,0)$ (left)  and 1-or-4-in-5 satisfiability $A=(0,1,0,0,1,0)$ (right) on a random factor graph with  $M$ constraints and average degree $\overline{\ell}$ (right), plotted versus $\overline{\ell}$ .
For comparison, the performance of simple belief-guided decimation process is shown. The vertical lines show the clustering and satisfiability thresholds.
}
\end{figure}
The performance of this algorithm is shown for 
1-in-4 satisfiability $A=(0,1,0,0,0)$ and 1-or-4-in-5 satisfiability $A=(0,1,0,0,1,0)$ in Fig.\ref{fig:decimation}. 
For 1-in-4 satisfiability, data with randomization is presented:
instead of fixing the most polarized variable or pair, we fix a variable or pair
randomly chosen
among a fixed number (here we adopt 8) of most polarized variables/pairs.
The figure also shows the two important thresholds for these problems, which are  values of the average degree (a measure of the number of constraints) separating
qualitatively distinct phases.
The probability that a satisfying assignment exists drops from 1 to 0 at the `satisfiability threshold' $\ell_{s}$ in the large factor graph limit. Between the `clustering threshold'
(\cite{ZM} and the satisfiability threshold, although the satisfying assignments still exist with probability one, it is very difficult to find one by the algorithms known so far, because of the splitting of the set of solutions into clusters.
In both LOPs the performance is improved compared to the
simple belief-guided decimation employed in \cite{ZM}.
Especially for 1-or-4-in-5, the present algorithm works well above the clustering threshold, a region of $\overline{\ell}$ where all known algorithms are reported to perform poorly \cite{ZM}.

\section{Conclusion and Discussion}
\label{sec:conclusion}
We have shown how to find satisfying assignments for locked occupation
problems based on the measurement of correlation among variables.
This is in contrast with the conventional method which is guided by 1-variable
marginals only.  Since flipping a variable in a LOP forces another variable
far apart to be flipped, the performance of the algorithm is 
improved when we take the correlations into account.

We have calculated correlations with the susceptibility propagation.
In this method, the correlations between variables which ar efar apart can be calculated as well as between those which are neighbors. Namely, the convergence property is controlled by a single matrix $\mathcal{M}$. 

The susceptibility propagation, however, requires more computational time and memory resource than the simple belief propagation.
Therefore, as the problem becomes larger, we face a (polynomial) increase of computation time.
The truncation introduced in subsection\ref{sec:decay} might give a remedy since it reduces by a factor of $N$ the computation time as well as the memory use.
The decay of correlation suggests that this is a reasonable approximation.
We have performed preliminary experiments to find the performance of this approximate algorithm. As expected,  it behaves similarly to that without truncation, the performance being only slightly degraded.

\section*{Acknowledgment}
S.H. was supported by Ryukoku University Research Fellowship (2008).

\bibliographystyle{apsrev}  
\bibliography{bp}

\end{document}